\documentclass[12pt]{article}

\usepackage{amsmath,amssymb,amsfonts,mathtools}
\usepackage{bm,bbm}
\usepackage{booktabs}
\usepackage{color}
\usepackage[dvips,letterpaper]{geometry}
\usepackage{epsfig}
\usepackage{graphicx,psfrag,epsf}
\usepackage{indentfirst}
\usepackage{lineno}
\usepackage{natbib}
\usepackage{setspace}
\usepackage{subfigure}
\usepackage{times}
\usepackage{url}
\usepackage{verbatim}
\usepackage{multirow}
\usepackage[table,xcdraw]{xcolor}
\usepackage{subeqnarray}
\usepackage{breqn} 
\usepackage{here}
\usepackage{adjustbox}

\usepackage[draft,authormarkuptext=name]{changes}
\usepackage{lipsum}% <- For dummy text
\definechangesauthor[name={Santos-Neto}, color=cyan]{sn} % id Manoel
\definechangesauthor[name={Victor Leiva}, color=blue]{vl} % id Manoel
\definechangesauthor[name={Helton Saulo}, color=orange]{hs} % id Manoel
\definechangesauthor[name={Mário},color=green]{ms} % id Manoel

\setremarkmarkup{(#2)}
\input cyracc.def

\def \iid  {\stackrel{\textrm{\tiny IID}}{\sim}}

\title{\textbf{\LARGE A new mixture-based fixed-effect model for a biometrical case-study related to immunogenecity with highly censored data}}

 \author{\normalsize
 \textbf{M\'{a}rio F. Desousa}$^{1,6}$, \textbf{Helton Saulo}$^{2}$\,,  
 \textbf{Manoel Santos-Neto}$^{3,4}$\,, \textbf{V\'{i}ctor Leiva}$^{5,6}$\thanks{V\'{i}ctor Leiva, Email: victorleivasanchez@gmail.com, URL: www.victorleiva.cl}\\
 {\footnotesize $^{1}$Department of Statistics, Universidade Estadual de Campinas, Brazil}\\[-0.2cm]
 {\footnotesize $^{2}$Department of Statistics, Universidade de Bras\'ilia, Brazil}\\[-0.2cm]
 {\footnotesize $^{3}$Department of Statistics, Universidade Federal de S\~ao Carlos, Brazil }\\[-0.2cm]
{\footnotesize $^{4}$Department of Statistics, Universidade Federal de Campina Grande, Brazil }\\[-0.2cm]
 {\footnotesize $^{5}$School of Industrial Engineering, Pontifica Universidad Cat\'{o}lica de Valpara\'{i}so, Chile }\\[-0.2cm]
 {\footnotesize $^{6}$Faculty of Management, Accounting and Economics, Universidade Federal de Goi\'as, Brazil}}  %[-0.15cm]
\date{}
 
\begin{document}
%\listofchanges

\maketitle
\vspace{-0.75cm}
\begin{abstract}
We propose a new continuous-discrete mixture regression model which is useful for describing highly censored data. We motivate our investigation based on a case-study in biometry related to measles vaccines in Haiti. In this case-study, the neutralization antibody level is explained by the type of vaccine used, level of the dosage and gender of the patient. This mixture model allows us to account for excess of censored observations and consists of the Birnbaum-Saunders and Bernoulli distributions. These distributions describe the antibody level and the point mass of the censoring observations. We estimate the model parameters with the maximum likelihood method. Numerical evaluation of the model is performed by Monte Carlo simulations and by an illustration with biometrical data, both of which show its good performance and its potential applications. 

\vspace{-0.25cm}
\end{abstract}

\section{Bibliographical review and motivating example}%\label{int}

In this section, we provide an introduction to the topic accompanied by a state of art about studies linked to the present investigation. In addition,  a motivating example from biometry is presented to justify the development of the proposed methodology.

%\newpage
\subsection{Introduction}%\label{sec:introd}

A frequently studied topic in survival models is the censored data analysis. Particularly, tobit models are used to estimate parameters of interest when censored data are present; see \cite{l:96} and \cite{km:03} for details on tobit models as well as censored and truncated data, respectively. However, we detect three problems regarding the standard tobit model. First, it has a strong assumption which is the normality (and therefore symmetry) for the model error. Second, the standard tobit model does not cover situations of extreme heaviness for the censored part of the distribution. Third, tobit models does not take into account the lower detection limit (LDL) and the possible existence of some observations below this LDL. This situation is present in studies of immunogenecity related to measles vaccine data; see details in Section \ref{sec:motiexa}. Therefore, first, as it is well-known, ignoring the effect of asymmetry can be harmful and lead to significantly biased estimates. Then, some flexible tobit models, in terms of kurtosis and asymmetry, are been introduced by \cite{martinezetal:13b}, \cite{rochaetal:15}, \cite{bgls:18} and \cite{dsls:16}, but these tobit models do not solve second and third problems.

A two-part model proposed by \cite{cragg:71} solves the problem of a large number of censored observations. That model considers the possibility of having observations from the assumed distribution for data with positive support (part 1) and from a point mass distribution (part 2). In the model, the log-normal (LN) distribution was considered for the positive response variable. However, this model based on the LN distribution does not consider 
the existence of both an LDL and some observations at or below this LDL. \cite{mh:95} proposed a 
generalization of the two-part model, named Bernoulli/LN model, by considering the possibility of limited responses resulting from interval censoring associated with the positive support distribution. In the generalized two-part model, any value above the LDL obligatorily comes from the LN distribution, whereas a censored value may come from either the point mass distribution or the LN distribution.

The Birnbaum-Saunders (BS) distribution is unimodal, positively skewed and has a close relation with the normal distribution, such as the LN distribution; see \cite{bs:69a}, \cite{jkb:95}  and \cite{l:15}. The BS distribution has two parameters related to its shape and scale, where the latter one is also its median. Thus, the BS distribution can be seen as an analogue to the normal distribution, but in an  asymmetric setting, where the median is generally considered to be a better measure of central tendency than the mean. The BS distribution has been applied to model business, engineering, environmental and industry data, which
have been conducted by international, transdisciplinary groups of researchers. Some of its recent applications are attributed to \citet{slzm:13}, \citet{sclb:14,sclb:16}, \citet{lmrs:15,lfgl:15,lscb:15,lrsv:17}, 
\citet{wl:15}, \citet{mlcv:16,mlcl:17} and \citet{garcialru:18,garcialua:18}. The BS distribution has shown to be a good alternative to describe medical data in the works by \citet{leaolst:16,leaolst:17}. However and more relevantly, its adequacy to model medical data was justified in the recently work by \citet{leaolst:18} using mathematical arguments based on a conceptual analogy between material fatigue and medical settings.

The main objective of this paper is to propose a fixed-effect (regression) model for left-censored data based on the mixture between the BS and Bernoulli distributions, that is, a skew positive continuous distribution and a point mass distribution located below the  LDL. The proposed model extends the Bernoulli/LN model to the BS case.  The secondary objectives of this paper are: (i) to develop inference for the Bernoulli/BS model based on the maximum likelihood (ML) method; (ii) to perform a Monte Carlo (MC) simulation study to evaluate the performance of the ML estimators; and (iii) to carry out an application of the proposed model to an immunogenecity 
study of measles vaccine in Haiti. Thus, the Bernoulli/BS model appears as a new alternative to describe censored data. In order to motivate our research, we describe the following example with medical data related to immunogenecity.

\subsection{Motivating example in biometry}\label{sec:motiexa}

Determination of antibody concentration by quantitative assays is an important topic of research. In such a topic, often there is a concentration value ($V_c$) below which an exact measurement cannot be obtained, regardless of the employed technique. However, this antibody concentration value $V_c$ is a function of the associated assay. When left-censoring is present in data from an assay, the LDL can be used to substitute the value of the censored observation by using $V_c$. In special, this substitution is applied to immunogenecity studies where data are often censored. Then, tobit models could be used to estimate the parameters of interest. However, statistical modeling for data analysis of this type are not yet fully disseminated and the
topic is still an object of discussion.

\begin{figure}[!ht]
\centering
\psfrag{R}[c]{\scriptsize{empirical quantile}}
\psfrag{Q}[c]{\scriptsize{theoretical quantile}}
\psfrag{0}[c][c]{\scriptsize{0}}
\psfrag{1}[c][c]{\scriptsize{1}}
\psfrag{2}[c][c]{\scriptsize{2}}
\psfrag{3}[c][c]{\scriptsize{3}}
\psfrag{4}[c][c]{\scriptsize{4}}
\psfrag{5}[c][c]{\scriptsize{5}}
\psfrag{6}[c][c]{\scriptsize{6}}
\psfrag{7}[c][c]{\scriptsize{7}}
\psfrag{8}[c][c]{\scriptsize{8}}
\psfrag{9}[c][c]{\scriptsize{9}}
\psfrag{10}[c][c]{\scriptsize{10}}
\psfrag{15}[c][c]{\scriptsize{15}}
\psfrag{-0.002}[c][c]{\scriptsize{$-$.002}}
\psfrag{0.000}[c][c]{\scriptsize{.0}}
\psfrag{0.002}[c][c]{\scriptsize{.002}}
\psfrag{0.004}[c][c]{\scriptsize{.004}}
\psfrag{0.006}[c][c]{\scriptsize{.006}}
\psfrag{0.008}[c][c]{\scriptsize{.008}}
\psfrag{0.010}[c][c]{\scriptsize{.010}}
\psfrag{0.0}[c][c]{\scriptsize{0.0}}
\psfrag{0.1}[c][c]{\scriptsize{0.1}}
\psfrag{0.2}[c][c]{\scriptsize{0.2}}
\psfrag{0.3}[c][c]{\scriptsize{0.3}}
\psfrag{0.4}[c][c]{\scriptsize{0.4}}
\psfrag{0.5}[c][c]{\scriptsize{0.5}}
\psfrag{0.6}[c][c]{\scriptsize{0.6}}
\psfrag{0.7}[c][c]{\scriptsize{0.7}}
\psfrag{0.8}[c][c]{\scriptsize{0.8}}
\psfrag{1.0}[c][c]{\scriptsize{1.0}}
\psfrag{0}[c][c]{\scriptsize{0}}
\psfrag{50}[c][c]{\scriptsize{50}}
\psfrag{100}[c][c]{\scriptsize{100}}
\psfrag{150}[c][c]{\scriptsize{150}}
\psfrag{200}[c][c]{\scriptsize{200}}
\psfrag{250}[c][c]{\scriptsize{250}}
\psfrag{300}[c][c]{\scriptsize{300}}
\psfrag{328}[c]{\scriptsize{$328$}}
\psfrag{330}[l]{\scriptsize{$330$}}
\psfrag{in}[c]{\scriptsize{index}}
\psfrag{GCDn}{\scriptsize{GCD($\bm\theta$)}}
\subfigure[tobit-normal]{\includegraphics[height=7.5cm,width=7.5cm]{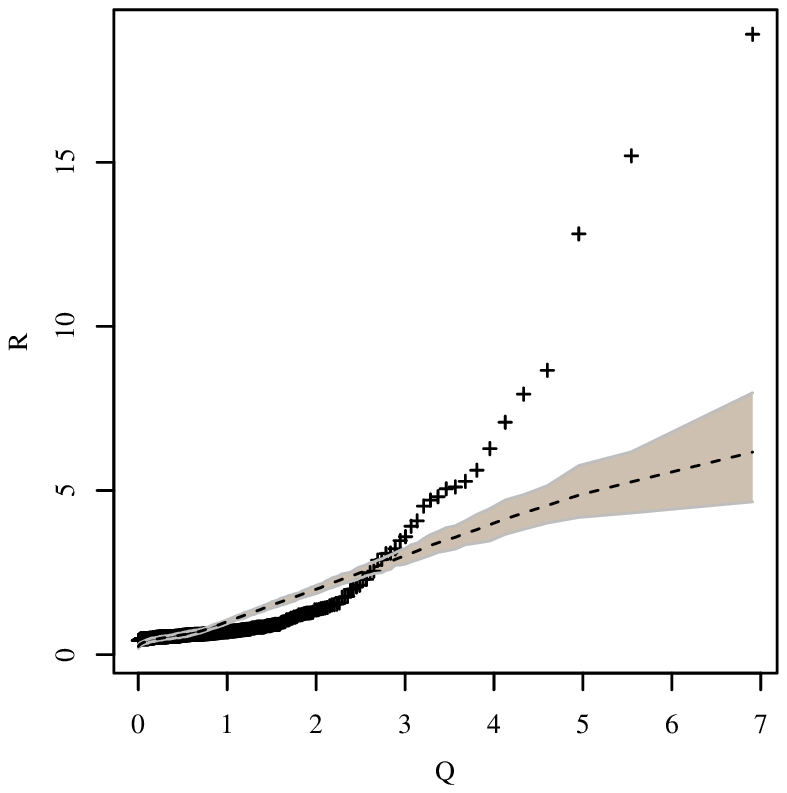}}
\subfigure[tobit-t]{\includegraphics[height=7.5cm,width=7.5cm]{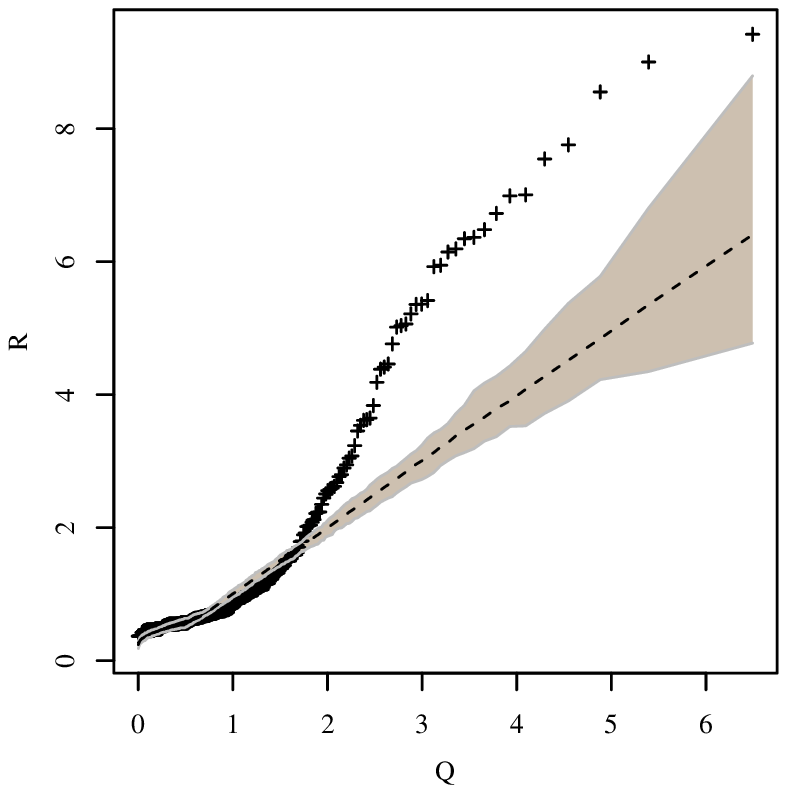}}
%\vspace{-0.25cm}
\caption{\small QQ plot and its envelope for the GCS residuals with the indicated model using vaccine data.}%
\label{fig:qqplots1}
\end{figure}

The motivation for our study came from a real-world medical data set provided by \cite{mh:95} about a safety and immunogenecity study related to measles vaccines in Haiti. In this case-study, the variable of interest (response) is the neutralization antibody level, whereas the following explanatory variables (covariates) were considered to explain this response: (i) EZ is the type of vaccine used (0 if Schwartz and 1 if Edmonston-Zagreb); (ii) HI is the level of the dosage  (0 if medium and 1 if high); and (iii) FEM is the gender (0 for male and 1 for female). Then, a regression model could be used to describe the relationship between the response and covariates. However, the response was observed in 330 children at 12 months of age, of which 86 (26.1\%) cases had a neutralization antibody level below the LDL and then such levels were recorded with the corresponding $V_c$. Note that in this study the LDL was $V_c = 0.1$, in international units, or $-$2.306 in logarithm scale. Therefore, a first natural approach for describing these data can be the tobit model based  on the normal distribution (tobit-normal); see details in \cite{bggl:10} and Section \ref{tobit:models}. Figure \ref{fig:qqplots1}(a) displays the QQ plot with simulated envelopes of the generalized Cox-Snell (GCS) residual based on the tobit-normal model; see Section \ref{sec:3.4} for details about this residual. This figure shows that the GCS residuals provide a bad performance of this first model fitted to measles vaccines data. We observe from this figure that the bad fitting is detected mainly in the right tail. Therefore, a tobit model based on heavy-tailed distribution, such as the t distribution, might improve the fitting. Then, we consider the tobit-t model for this possible improvement. Figure \ref{fig:qqplots1}(b) presents a similar plot to that Figure \ref{fig:qqplots1}(a) but now based on the tobit-t model; see Section \ref{tobit:models} for details of this model. Once again, the residual plots shows a bad performance now of the tobit-t model, so that we deduct the problem was not due to the right tail but to a posible asymmetry, because we have evaluated only tobit-symmetric models. Next, an exploratory data analysis is conducted to detect asymmetry and kurtosis.

\newpage
\vspace{0.25cm}

\vspace{0.25cm}

\begin{figure}[h!]
\centering
\psfrag{0.000}[c][c]{\scriptsize{.0}}
\psfrag{0.001}[c][c]{\scriptsize{.001}}
\psfrag{0.002}[c][c]{\scriptsize{.002}}
\psfrag{0.003}[c][c]{\scriptsize{.003}}
\psfrag{0.004}[c][c]{\scriptsize{.004}}
\psfrag{0.005}[c][c]{\scriptsize{.005}}
\psfrag{0.006}[c][c]{\scriptsize{.006}}
\psfrag{0.0}[c][c]{\scriptsize{0.0}}
\psfrag{0.1}[c][c]{\scriptsize{0.1}}
\psfrag{0.2}[c][c]{\scriptsize{0.2}}
\psfrag{0.3}[c][c]{\scriptsize{0.3}}
\psfrag{0.4}[c][c]{\scriptsize{0.4}}
\psfrag{0.5}[c][c]{\scriptsize{0.5}}
\psfrag{0.6}[c][c]{\scriptsize{0.6}}
\psfrag{0.7}[c][c]{\scriptsize{0.7}}
\psfrag{0.8}[c][c]{\scriptsize{0.8}}
\psfrag{1.0}[c][c]{\scriptsize{1.0}}
\psfrag{5}[c][c]{\scriptsize{5}}
\psfrag{10}[c][c]{\scriptsize{10}}
\psfrag{15}[c][c]{\scriptsize{15}}
\psfrag{20}[c][c]{\scriptsize{20}}
\psfrag{0}[c][c]{\scriptsize{0}}
\psfrag{200}[c][c]{\scriptsize{200}}
\psfrag{400}[c][c]{\scriptsize{400}}
\psfrag{600}[c][c]{\scriptsize{600}}
\psfrag{800}[c][c]{\scriptsize{800}}
\psfrag{1000}[c][c]{\scriptsize{1000}}
\psfrag{dp}[c][c]{\scriptsize{$y$}}
\psfrag{de}[c][c]{\scriptsize{PDF}}
\psfrag{ab}[c][c]{\scriptsize{$W_{n}(k/n)$}}
\psfrag{in}[c][c]{\scriptsize{$k/n$}}
\psfrag{aa}[c][c]{\scriptsize{usual boxplot}}
\psfrag{ad}[c][c]{\scriptsize{adjusted boxplot}}
\subfigure[]{\includegraphics[height=7cm,width=7.25cm]{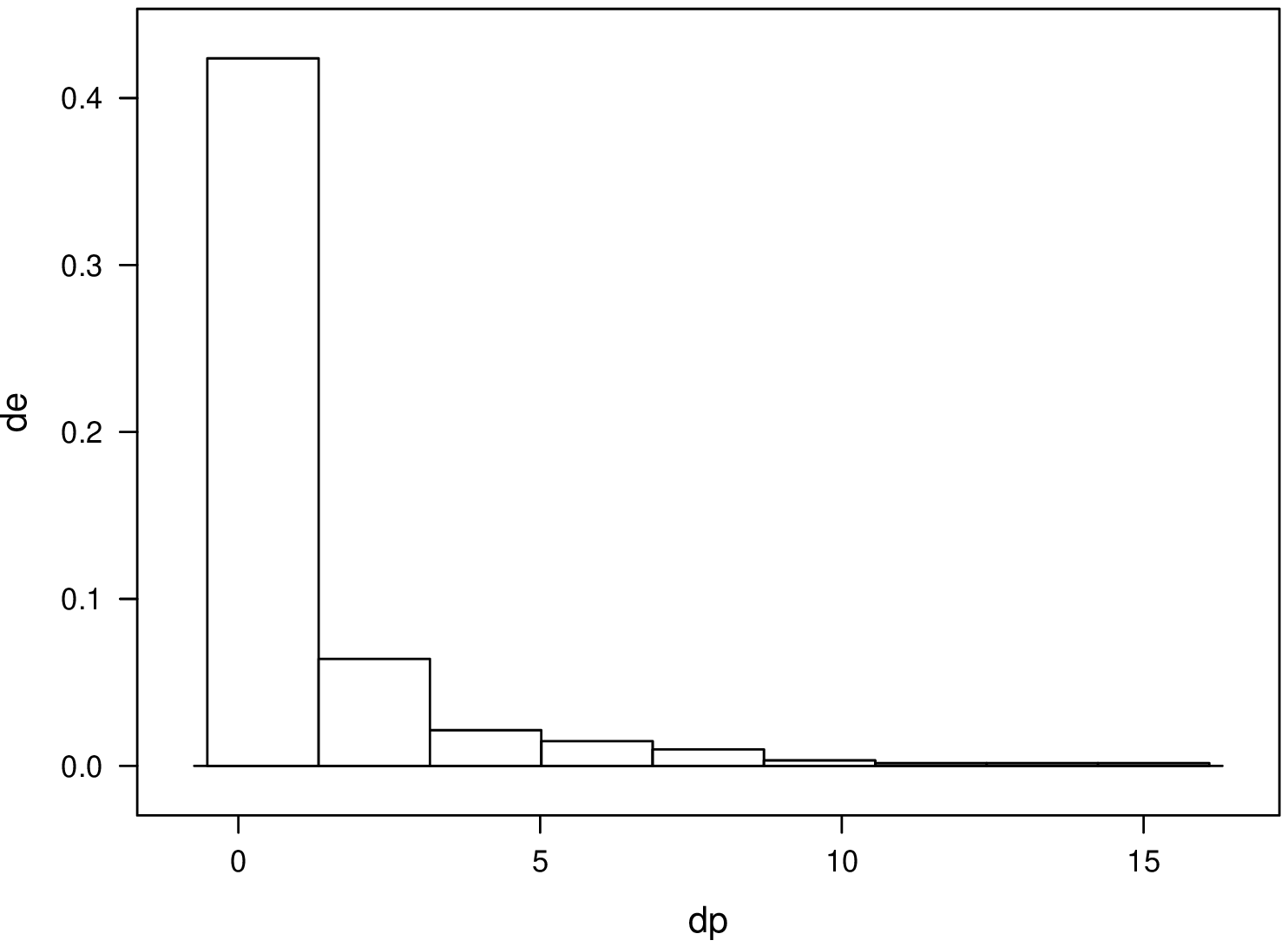}}
\subfigure[]{\includegraphics[height=7cm,width=7.25cm]{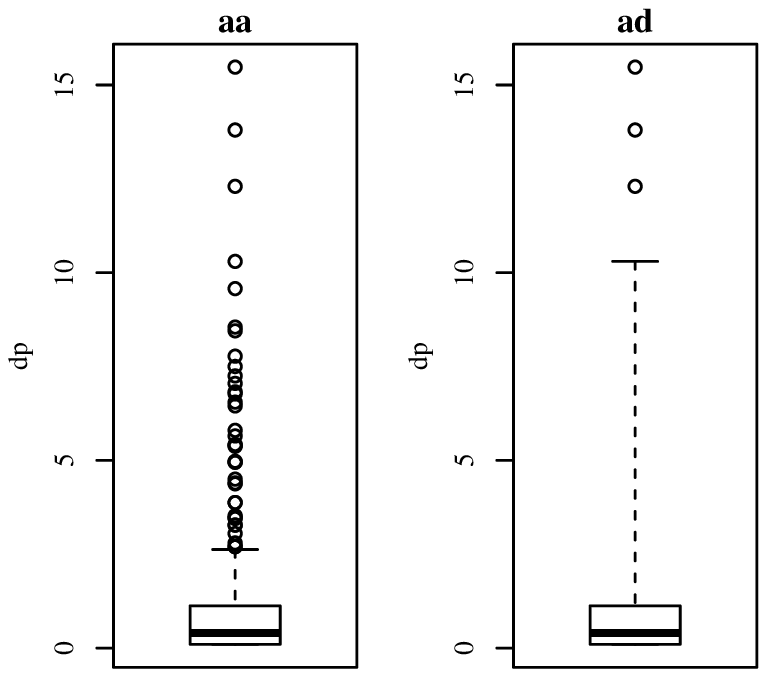}}
\vspace{-0.25cm}
\caption{\small Histogram  (a) and boxplots (b) for vaccine data.}
  \label{fig:2exploratory}
\end{figure}

\begin{figure}[!ht]
\centering
\psfrag{R}[c]{\scriptsize{empirical quantile}}
\psfrag{Q}[c]{\scriptsize{theoretical quantile}}
\psfrag{0}[c][c]{\scriptsize{0}}
\psfrag{1}[c][c]{\scriptsize{1}}
\psfrag{2}[c][c]{\scriptsize{2}}
\psfrag{3}[c][c]{\scriptsize{3}}
\psfrag{4}[c][c]{\scriptsize{4}}
\psfrag{5}[c][c]{\scriptsize{5}}
\psfrag{6}[c][c]{\scriptsize{6}}
\psfrag{7}[c][c]{\scriptsize{7}}
\psfrag{8}[c][c]{\scriptsize{8}}
\psfrag{9}[c][c]{\scriptsize{9}}
\psfrag{10}[c][c]{\scriptsize{10}}
\psfrag{15}[c][c]{\scriptsize{15}}
\psfrag{-0.002}[c][c]{\scriptsize{$-$.002}}
\psfrag{0.000}[c][c]{\scriptsize{.0}}
\psfrag{0.002}[c][c]{\scriptsize{.002}}
\psfrag{0.004}[c][c]{\scriptsize{.004}}
\psfrag{0.006}[c][c]{\scriptsize{.006}}
\psfrag{0.008}[c][c]{\scriptsize{.008}}
\psfrag{0.010}[c][c]{\scriptsize{.010}}
\psfrag{0.0}[c][c]{\scriptsize{0.0}}
\psfrag{0.1}[c][c]{\scriptsize{0.1}}
\psfrag{0.2}[c][c]{\scriptsize{0.2}}
\psfrag{0.3}[c][c]{\scriptsize{0.3}}
\psfrag{0.4}[c][c]{\scriptsize{0.4}}
\psfrag{0.5}[c][c]{\scriptsize{0.5}}
\psfrag{0.6}[c][c]{\scriptsize{0.6}}
\psfrag{0.7}[c][c]{\scriptsize{0.7}}
\psfrag{0.8}[c][c]{\scriptsize{0.8}}
\psfrag{1.0}[c][c]{\scriptsize{1.0}}
\psfrag{0}[c][c]{\scriptsize{0}}
\psfrag{50}[c][c]{\scriptsize{50}}
\psfrag{100}[c][c]{\scriptsize{100}}
\psfrag{150}[c][c]{\scriptsize{150}}
\psfrag{200}[c][c]{\scriptsize{200}}
\psfrag{250}[c][c]{\scriptsize{250}}
\psfrag{300}[c][c]{\scriptsize{300}}
\psfrag{328}[c]{\scriptsize{$328$}}
\psfrag{330}[l]{\scriptsize{$330$}}
\psfrag{in}[c]{\scriptsize{index}}
\psfrag{GCDn}{\scriptsize{GCD($\bm\theta$)}}
\subfigure[tobit-LN]{\includegraphics[height=7cm,width=7.25cm]{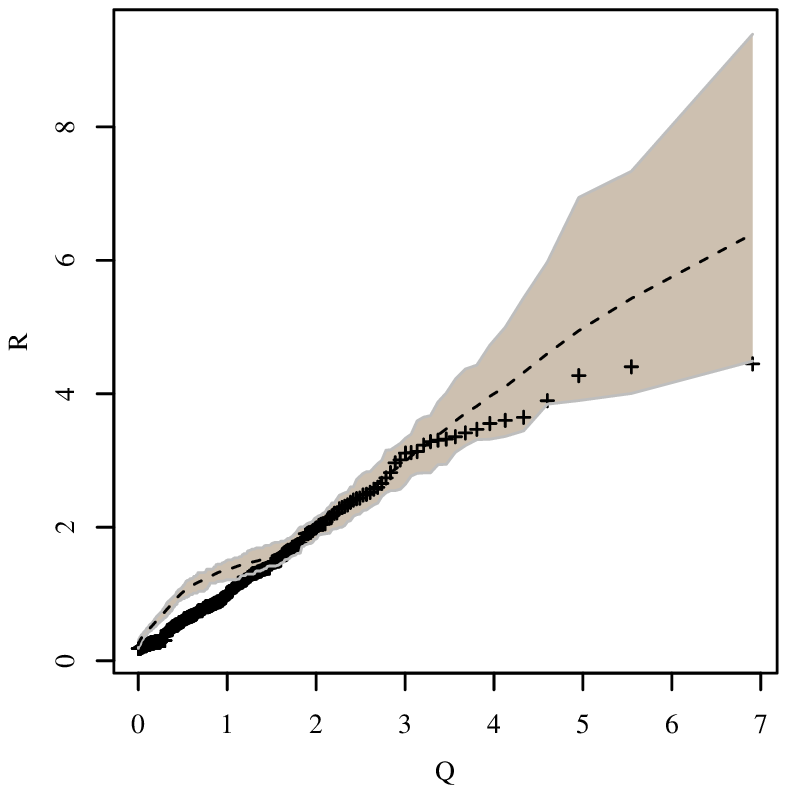}}
\subfigure[tobit-BS]{\includegraphics[height=7cm,width=7.25cm]{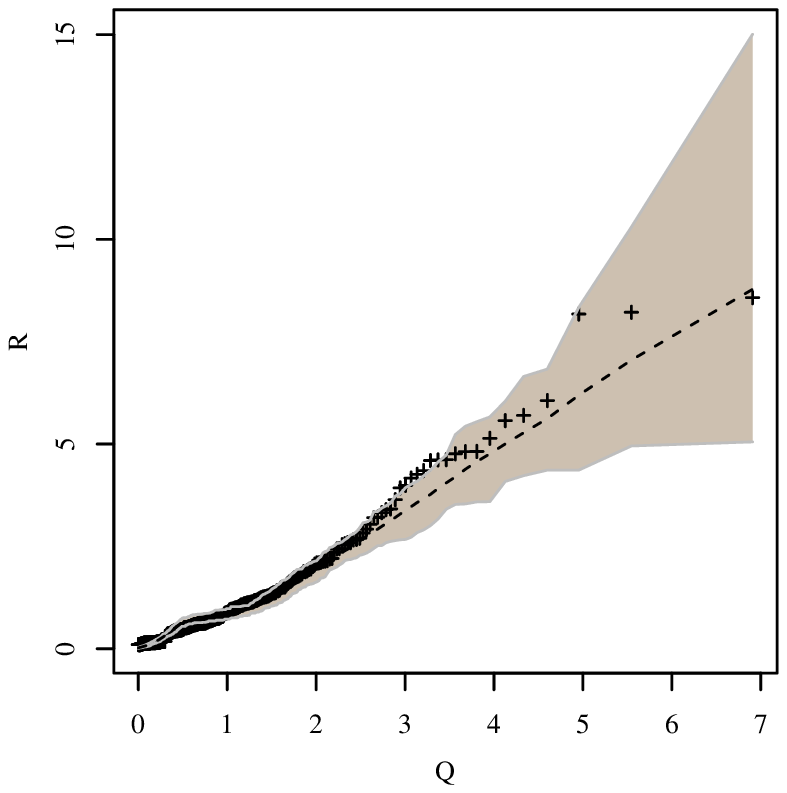}}
\vspace{-0.25cm}
\caption{\small QQ plot and its envelope for GCS residual with the indicated model using vaccine data.}%
\label{fig:qqplots2}
\end{figure}

Table~\ref{tab:2estdesc} provides descriptive statistics for the measles vaccine data set, including minimum, maximum, median, mean, standard deviation (SD) and coefficients of variation (CV), skewness (CS) and kurtosis (CK). The CK and CS indicate the positive skew nature and high kurtosis level of the data distribution.  Figure~\ref{fig:2exploratory} shows the histogram and boxplots for the measles vaccine data.  From this figure, note that the skewed nature reported in Table~\ref{tab:2estdesc} is confirmed by the histogram of Figure~\ref{fig:2exploratory}(a). Note that some outliers considered by the usual boxplot presented in Figure~\ref{fig:2exploratory}(b) are not outliers when we consider the adjusted boxplot; see details on this latter boxplot in \cite{rctrsvkm:16}. Then, under asymmetry, a heavy-tailed distribution is not needed, but rather a positive skew distribution. Therefore, we consider tobit-LN and tobit-BS models; see details about these models in \cite{dsls:16}.  Figure \ref{fig:qqplots2}(a) displays the QQ plot with simulated envelopes of the GCS residual based on the tobit-LN and tobit-BS models. This figure shows a better performance of the tobit-LN model in relation to the tobit-normal and tobit-t models. However, the tobit-LN model is still inappropriate, but the tobit-BS model seems to be appropriate, although some fitting problems are detected at the tails possibly due to the extreme percentage of censoring.

\vspace{0.25cm}
\begin{table}[h!]
%\small
\centering
\caption{\small Descriptive statistics for vaccine data.}
\label{tab:2estdesc}
\begin{tabular}{ccccccccc} \toprule
$n$ & Min   &   Max    &   Mean   & Median    & SD     &  CV       &  CS   &  CK   \\
\midrule
330 & {0.10}    &  15.47   &  1.20    & {0.40}     & 2.10   & 174.74\%  &  3.46  & 14.37 \\
\bottomrule
\end{tabular}
\end{table}

\vspace{0.25cm}

\newpage
In summary, it is necessary to consider a model to analyze measles vaccine data, which have censoring and asymmetry. However,  the model to be postulated must consider an LDL and high censoring. As mentioned, tobit models do not consider this limit and  its omission can distort the results obtained from the corresponding analysis. Therefore, this example serves as a motivation to formulate a model which allows us to describe high censoring, asymmetry and an LDL. The model to be formulated should be based on a distribution with theoretical arguments useful in biometry (as the BS distribution), to account for excess of censored observations and to estimate a proportion that determines the contribution of the point mass distribution.

\subsection{Organization of the paper}%\label{sec_me}

The rest of the paper proceeds as follows. Section \ref{sec:2} provides a background of the BS distribution and its logarithmic transformation, as well as of tobit models and mixture models for left-censored data. In Section \ref{sec:3}, we formulate the Bernoulli/BS model along with inference and estimation based on the ML method. In Section \ref{sec:4}, the model is evaluated through MC simulations and illustrated with biometrical data related to measles vaccines in Haiti. Some concluding remarks and possible future research are mentioned in Section \ref{sec:6}.

\section{Preliminaries}\label{sec:2}

In this section, we present a background related to (i) the BS distribution and its logarithmic transformation; (ii) tobit models; and iii) mixture models for left-censored data.

\subsection{BS and log-BS distributions}

Let $T$ be a random variable with BS distribution of shape ($\alpha$) and scale ($\sigma$) parameters, denoted it by $T\sim\textrm{BS}(\alpha,\sigma)$. Then,   the probability density function (PDF) of $T$ is expressed as
\begin{equation}\label{eqnew:1}
f_T(t;\alpha,\sigma)
 = \frac{1}{2\alpha}
\left(\sqrt{{1}/{\sigma t}}+\sqrt{{\sigma}/{t^{\frac{3}{2}}}}\right)
\phi\left(\frac{1}{\alpha}\left(\sqrt{{t}/{\sigma}}-\sqrt{{\sigma}/{t}}\right)\right), \, t>0, \alpha>0, \sigma>0,
\end{equation}
where $\phi$ is the standard normal PDF. When covariates (${\bm X}_{i}$) are added in a statistical modeling based on the BS distribution with PDF as given in \eqref{eqnew:1}, the relation between the response variable ($T_i$) and the observed values (${\bm x}_{i}$) of these covariates is often non-linear with an exponential structure, as usual in asymmetric data; see \cite{mlcv:16}. Then, in order to formulate fixed-effect models under a BS setting, one transforms the exponential regression structure to a linear one of standard type as
\begin{equation}\label{eqnew:2}
Y_i = \bm{x}_{i}^{\top}{\bm\beta} + \varepsilon_{i}, \quad i = 1, \ldots,n,
\end{equation}
where $Y_i = \log(T_i)$, ${\bm x}_{i}^{\top}=(x_{i1},x_{i2},\ldots,x_{ip})$ is the $i$th observation on a set of $p$ independent covariates ${\bm X}_{i}$,  ${\bm\beta}^{\top}=(\beta_{1},\beta_{2},\ldots,\beta_{p})$ is a vector of fixed effect parameters to be estimated, $\varepsilon_{i} $ is the error term of the model. Note that $\varepsilon_{i}$ defined in \eqref{eqnew:2} corresponds to $\varepsilon_{i} = \log(\delta_{i})$, where $\delta_{i} \sim\textrm{BS}(\alpha,1)$. Then, this modeling framework needs the use of a logarithmic version of the BS distribution (log-BS) defined as follows. A random variable $Y$ has a log-BS distribution with shape ($\alpha>0$) and location ($\mu \in \mathbb{R}$) parameters, denoted it as $\textrm{log-BS}(\alpha,\mu)$, if and only if $Z=(2/\alpha)\textrm{sinh}({(Y-\mu)}/{2})\sim \textrm{N}(0,1)$, where $\mu=\log(\sigma)$. Then, the cumulative distribution function (CDF) of $Y$ is given by
\begin{equation}\label{eq:2CDFlogBS}
F_Y(y; \alpha, \mu)=\Phi\left(\frac{2}{\alpha}\textrm{sinh}\left(\frac{y-\mu}{2}\right)\right),\quad
y \in \mathbb{R}, \mu \in \mathbb{R}, \alpha>0.
\end{equation}
Consequently, from \eqref{eq:2CDFlogBS}, the PDF of $Y$ is defined as
\begin{equation}\label{eq:2PDFlogBS}
f_Y(y; \alpha, \mu)=\frac{1}{\alpha\sqrt{2\pi}}\textrm{ cosh}\left(\frac{y-\mu}{2}\right)
\textrm{exp}\left( -\frac{2}{\alpha^2}\textrm{sinh}^2\left(\frac{y-\mu}{2}\right)\right), \, y \in \mathbb{R}, \mu \in \mathbb{R}, \alpha>0,
\end{equation}
whereas the logarithm of the PDF given in \eqref{eq:2PDFlogBS} is expressed as
$$
\log(f_Y(y; \alpha, \mu)) = -\log(2) - \frac{\log(2\pi)}{2} + \log\left( \frac{2}{\alpha}\textrm{cosh}\left(\frac{y-\mu}{2}\right)\right) -\frac{2}{\alpha^2}\left(\textrm{sinh}\left( \frac{y-\mu}{2}\right)\right)^2, 
$$
for $y \in \mathbb{R}$, which is useful for several purpose, such as in likelihood-based methods.

\subsection{Tobit models}\label{tobit:models}

Consider a sample of size $n$, ${\bm Y}=(Y_1,\ldots,Y_m,Y_{m+1},\ldots,Y_n)^{\top}$ namely, composed by independent (IND) random variables 
but not necessarily independent identically distributed (IID). Consider also that this sample includes $m$ censored data to the left 
and $n-m$ complete or uncensored data. 

\begin{figure}[!ht]
\centering
\vspace{-0.25cm}
\psfrag{Y}[r][c][0.5]{\large{$Y_i, Y^*_i$}}
\psfrag{A}[l][c][0.5]{\hspace{-1cm}\large{$Y_i$ is equal to $\xi$}}
\psfrag{B}[l][c][0.5]{\large{$Y^*_i$ is greater than $\xi$}}
\psfrag{V}[l][c][0.5]{\large{$X_{i}$}}
\psfrag{E}[l][c][0.5]{\large{$\xi$}}
\psfrag{C}[l][c][0.5]{\large{censored observations when $Y^*_i$ is less than $\xi$}}
\includegraphics[height=5cm,width=7.5cm]{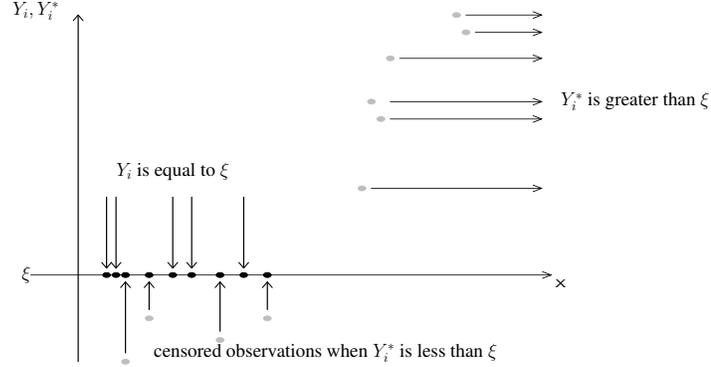}
\vspace{-0.25cm}
\caption{Illustration of the tobit model when $\xi=0$.}%
\label{fig:tobit}
\end{figure}

The tobit setting is formulated such that the $m$ censored data correspond 
to the values of $Y^{\ast}$ (censored response) less than or equal to a threshold point $\xi$ (censoring to the left), so that all of these data 
take the value $\xi$. The remaining $n-m$ uncensored data are related to values of $Y^{\ast}$ greater than $\xi$, which can be 
modeled by a linear regression structure. Then, the tobit model with 
censored response to the left can be written as
\begin{equation}\label{eq:21}
Y_{i} = \begin{cases}
\xi, & \; \textrm{if} \; Y_{i}^{\ast} \;  \leq \; \xi, \quad i = 1, \ldots,m;\\
\bm{x}_{i}^{\top}{\bm\beta} + \varepsilon_{i}, & \; \textrm{if} \; Y_{i}^{\ast} \; > \; \xi, \quad i = m+1, \ldots,n,
\end{cases}
\end{equation}
where $\varepsilon_{i} \iid \mathcal{F}$, $\bm{\beta}$ and $\bm{x}_{i}$ are as defined in \eqref{eqnew:2}. Table~\ref{tab:2} reports some tobit models according to the distribution $\mathcal{F}$ considered. Note that $\xi$ given in \eqref{eq:21} is a prefixed limiting value that makes the response of 
the regression model to be censored. Figure \ref{fig:tobit} provides an illustration of the tobit model presented in \eqref{eq:21} when $\xi=0$ with one covariate. Note that, when $Y^*_i$ is less than or equal to $\xi=0$, $Y_i$ is equal to $\xi=0$. The Tobit models use all of the information, including censoring.  

\begin{table}[H]
\centering
%\small
\caption{Different tobit models according to the error distribution and its bibliographical reference.}
\begin{tabular}{ll} 
\toprule
Distribution ($\mathcal{F}$)& Reference\\ 
\midrule
Normal& \cite{t:58} \\
LN& \cite{hsul:2008} \\
Student-$t$& \cite{bgls:18}  \\
BS& \cite{dsls:16} \\
 \bottomrule
\end{tabular}
\label{tab:2}
\end{table}

\subsection{Mixture models for highly censored data}

\cite{cragg:71} proposed a formulation to account for highly censored data described by
\begin{equation}\label{eq:cragg}
g(y_{i})=\pi_{i}\textrm{1}_{i}+(1-\pi_{i})f(y_{i})(1-\textrm{1}_{i}),
\end{equation}
where $0< \pi_{i} < 1$ is a weight factor that determines the contribution of the point mass distribution, $f$ is the LN PDF of a random variable $Y$ and $\textrm{1}_{i}$ is a function indicating the  value 0 if $y_{i}>V_c$ and 1 if $y_{i}\leq V_c$. Note that the PDF given in \eqref{eq:cragg} is not restricted to a specific statistical distribution, so that it can be switched by other models different to the LN one.  However,  the framework presented in \eqref{eq:cragg} does not consider the possibility of an LDL.

\cite{mh:95} extended the model proposed by \cite{cragg:71} to a generalized version of two-parts considering limiting responses coming from interval censoring.
This model incorporates an intermediary possibility 
that a censored value may be from either $f$ or from the point mass distribution. The PDF for the generalized two-part model is given by
\begin{equation}\label{eq:mh95}
g(y_{i})=\left(\pi_{i}+\left(1-\pi_{i}\right)F(V_c)\right)\textrm{1}_{i}+(1-\pi_{i})f(y_{i})(1-\textrm{1}_{i}),
\end{equation}
where $F$ is the CDF associated with the PDF $f$ and its corresponding CDF is obtained as
$$
G(y_i) = \begin{cases}
0, \; \textrm{if} \; y_i \leq 0;\\
\pi_i + (1-\pi_i) F(V_c), \; \textrm{if} \; 0 < y_i \leq V_c;\\
\pi_i + (1-\pi_i) F(V_c) + (1 - \pi_i) (F(y_i) - F(V_c)), \; \textrm{if}  \; y_i > V_c.
\end{cases}
$$
Note that a wide family of mixture models can be created by changing $f$ and the distribution associated with $\pi_{i}$ in \eqref{eq:mh95}. For example, the mixture estructure can be modeled with a dichotomous random variable $B$ with Bernoulli distribution of parameter $\textrm{P}(B=1)=\tau=1-\pi$. \cite{mbg:13} derived a Bernoulli/log-power-normal model.  Note that if $f$ corresponds to the $\textrm{N}(\bm{x}_{i}^{\top}{\bm\beta}, \sigma^2)$ distribution  and $\pi_{i}=0$, for $i=1,\ldots,n$, the formulation given in \eqref{eq:mh95} for the generalized  two-part model becomes the standard tobit model defined in \eqref{eq:21}. Nevertheless, as mentioned, the tobit setting is unable for modeling situations with excess of censored observations nor the presence of a LDL for observations censored below this LDL. 

\section{The Bernoulli/BS mixture model}\label{sec:3}

In this section, we formulate the new mixture model. Then, we estimate its parameters by the ML method. Details about inference for these parameters is also provided, as well as a residual analysis as diagnostic tool for model checking.

\subsection{Formulation}

We propose a mixture model between the Bernoulli and BS distributions (Bernoulli/BS) by  assuming that $f$ given in \eqref{eq:mh95} corresponds to the log-BS PDF defined in \eqref{eq:2PDFlogBS} and $\pi_i$ associated with the random variable $B$ following a Bernoulli distribution of parameter $1-\pi_i$. Then, the formulation defined in \eqref{eq:mh95} can be rewritten as follows
\begin{dmath}\label{eq:lk1}
g(y_{i})=\left(\pi_{i}+\left(1-\pi_{i}\right)\Phi\left(\zeta_{2_i}^{\textrm{\tiny{c}}}\right)\right)\textrm{1}_{i}+ (1-\pi_{i})
\left(
\frac{c_1}{\alpha}\cosh\left(\frac{y_{i}-\mu_{i}}{2}\right) \exp\left(-\frac{2}{\alpha^2}\textrm{sinh}^2\left(\frac{y_{i}-\mu_{i}}{2}\right)\right)\right)(1-\textrm{1}_{i}),
\nonumber
\end{dmath}
where $c_{1}=1/\sqrt{2\pi}$, $\mu_{i}={\bm x_{(1)_i}^{\top}{\bm\beta_{(1)}}}$, $\zeta^{\textrm{\tiny c}}_{2_i} 
= (2/\alpha) \sinh ( ({V_c} - {\bm x_{(1)_i}^{\top}{\bm\beta_{(1)}}})/2 )$, 
\begin{eqnarray}\label{eq:ID}
\textrm{1}_{i} =
\begin{cases}
 1, & \textrm{if} \; y\leq V_{c}; \\
 0, & \textrm{if} \; y > V_{c};
\end{cases}
\end{eqnarray}
$\Phi$ is the standard normal CDF, $\bm{x}_{(1)}$ is a vector of values for covariates associated with the log-response variable and $\bm{\beta}_{(1)}$ is the corresponding fixed-effect parameters.  For the parameter $\pi_i$ associated with the random variable $B$ earlier defined, we assume the logit link function 
\begin{equation}\label{eq:logitlink}
\textrm{logit}\left(\textrm{P}\left(B=1| \bm{x}_{i} \right)\right)=\bm{x}_{(2)_i}^{\top}{{\bm \beta_{(2)}}} \Longleftrightarrow
\tau_{i}=1-\pi_{i}=\frac{\exp(\bm{x}_{(2)_i}^{\top} \bm{\beta}_{(2)})}{1+\exp(\bm{x}_{(2)_i}^{\top}\bm{\beta}_{(2)})},
\end{equation}
where $\bm{x}_{(2)}$ are observed covariates related to fixed-effect parameters $\bm{\beta}_{(2)}$ of the link function.

\subsection{Estimation}

Combining expressions \eqref{eq:lk1} and \eqref{eq:logitlink}, we obtain the individual contribution to the likelihood function of parameter $\bm{\theta} = (\alpha, \bm{\beta}_{(1)}^\top, \bm{\beta}_{(2)}^\top)^\top$ of the mixture Bernoulli/BS model that is given by

\vspace{-0.25cm}
\begin{dmath}\label{eq:likelihood}
L_{i}({\bm\theta})=\left(1+\frac{\exp({\bm x}_{(2)_i}^{\top}{\bm\beta}_{(2)})}{1+\exp({\bm x}_{(2)_i}^{\top}{\bm\beta}_{(2)})}\left(\Phi\left(\zeta_{2_i}^{\textrm{\tiny{c}}}\right)-1\right)\right)^{\textrm{1}_i} \times 
\left(\frac{\exp({\bm x}_{(2)_i}^{\top}{\bm \beta}_{(2)})}{1+\exp({\bm x}_{(2)_i}^{\top}{\bm\beta}_{(2)})}\left(\frac{c_1}{\alpha}\cosh\left(\frac{y_{i}-\mu_{i}}{2}\right)\exp\left(-\frac{2}{\alpha^2}\textrm{sinh}^2\left(\frac{y_{i}-\mu_{i}}{2}\right)\right)\right)\right)^{1-\textrm{1}_i}, \nonumber
\end{dmath}

\vspace{-0.25cm}
\noindent where $c_{1}$, $\mu$ and $\zeta_{2_i}^{\textrm{\tiny{c}}}$ are as given in \eqref{eq:lk1} and $\textrm{1}_i$ in \eqref{eq:ID}.  

The log-likelihood function for $\bm{\theta} = (\alpha, \bm{\beta}_{(1)}^\top, \bm{\beta}_{(2)}^\top)^\top$ obtained by taking the logarithm of \eqref{eq:likelihood} is expressed as 
\begin{dmath}\label{eq:loglik}
\ell({\bm\theta})=-(n-m)\log(2)-(n-m)\frac{\log(2\pi)}{2}+\sum_{i=1}^{n}\textrm{1}_{i}\left(\log\left(1+\exp({\bm x}_{(2)_i}^{\top}{\bm\beta}_{(2)})\Phi\left(\zeta_{i2}^{\textrm{\tiny{c}}}\right)\right)-\log\left(1+\exp\left({\bm x}_{(2)_i}^{\top}{\bm\beta}_{(2)}\right)\right)\right)  + \sum_{i=1}^{n}(1-\textrm{1}_{i})\left(\bm{x}_{(2)_i}^{\top}{\bm\beta}_{(2)} +\log\left(\zeta_{1_i}\right)-\frac{1}{2}\zeta_{2_i}^{2}-\log\left(1+\exp({\bm x}_{(2)_i}^{\top}{\bm\beta}_{(2)})\right)\right),
\end{dmath}
where $\zeta_{2_i}^{\textrm{\tiny{c}}}$ is as given in \eqref{eq:lk1} and 
\begin{equation}
\label{eq:zetas}
\zeta_{1_i}  = \frac{2}{\alpha} \cosh \left(\frac{{y_i} - {\bm x_{(1)_i}^{\top}{\bm \beta_{(1)}}}}{2}\right),\quad
\zeta_{2_i}  = \frac{2}{\alpha} \sinh \left(\frac{{y_i} - {\bm x_{(1)_i}^{\top}{\bm \beta_{(1)}}}}{2}\right).
\end{equation}
To obtain the ML estimators, it is necessary to maximize the log-likelihood function given in \eqref{eq:loglik}. The corresponding score vector is defined as
$\dot{\bm \ell}= {\partial{\ell({\bm\theta})}}/{\partial{{\bm\theta}}} = (\dot{\ell}_{\alpha}, 
\dot{\bm \ell}_{{\bm \beta}_{(1)}}^{\top},\dot{\bm \ell}_{{\bm \beta}_{(2)}}^{\top})^\top$, which contains the first partial derivatives of \eqref{eq:loglik}, where
\begin{eqnarray}\label{eq:score}
\dot{\ell}_{\alpha} &=&
\begin{cases}
 -\dfrac{1}{\alpha}\left(\dfrac{\exp( \bm x_{(2)_i}\bm\beta_{(2)})\phi(\zeta_{2_i}^{\textrm{\tiny{c}}})\zeta_{2_i}^{\textrm{\tiny{c}}}}{1+\exp\left(\bm x_{(2)_i}\bm\beta_{(2)}\right)\Phi(\zeta_{2_i}^{\textrm{\tiny{c}}})}\right), & \, i=1,\ldots,m;\\
 \dfrac{1}{\alpha}(\zeta_{2_i}^{2}-1), & \, i=m+1,\ldots,n;
\end{cases}\nonumber \\ 
\dot{\bm \ell}_{{\bm\beta}{(1)}} &=&
\begin{cases}
 -\dfrac{\bm x_{(1)_i}}{2}\left(\dfrac{\exp(\bm x_{(2)_i}{\bm\beta_{(2)}})\phi(\zeta_{2_i}^{\textrm{\tiny{c}}})\zeta_{1_i}^{\textrm{\tiny{c}}}}{1+\exp(\bm x_{(2)_i}{\bm\beta_{(2)}})\Phi(\zeta_{2_i}^{\textrm{\tiny{c}}})}\right),  & \, i=1,\ldots,m;\\
\dfrac{\bm x_{(1)_i}}{2}\left(\zeta_{1_i}\zeta_{2_i}-\dfrac{\zeta_{2_i}}{\zeta_{1_i}}\right), & \, i=m+1,\ldots,n;
\end{cases}\\ \nonumber
\dot{\bm \ell}_{{\bm\beta}{(2)}} &=&
\begin{cases}
\bm x_{(2)_i}\left(\dfrac{\exp(\bm x_{(2)_i}{\bm \beta_{(2)}})\Phi(\zeta_{2_i}^{\textrm{\tiny{c}}})}{1+\exp(\bm x_{(2)_i}{\bm\beta_{(2)}})\Phi(\zeta_{2_i}^{\textrm{\tiny{c}}})}- \tau_{i} \right),  & \, i=1,\ldots,m; \\ \nonumber
\bm x_{(2)_i}(1- \tau_{i})), & \, i=m+1,\ldots,n, \nonumber
\end{cases}
\end{eqnarray}
where $\zeta_{1_i}$ and $\zeta_{2_i}$ are given by equation \eqref{eq:zetas}, with $\zeta_{1_i}^{\textrm{\tiny{c}}}$ and $\zeta_{2_i}^{\textrm{\tiny{c}}}$ being similarly given as in \eqref{eq:zetas} but using $\tau_i$ instead of $y_i$.  The ML estimator of ${\bm \theta}$ is obtained equating \eqref{eq:score} to zero . Note that the system of equations defined by $\dot{\ell}_{\alpha}=0$, $\dot{\bm \ell}_{{\bm\beta}{(1)}}=0$  and $\dot{\bm \ell}_{{\bm\beta}{(2)}}=0$ does not have an analytic solution. In this paper, we solve them by an iterative procedure for non-linear optimization known ad Broyden-Fletcher-Goldfarb-Shanno (BFGS) quasi-Newton method.
 
\subsection{Inference}

Considering that some regularity conditions discussed in \cite{ch:74} hold, the ML estimators $\widehat{\alpha}$, $\widehat{\bm\beta}_{(1)}$ and $\widehat{\bm\beta}_{(2)}$ are consistent and follow a multivariate normal joint asymptotic distribution with mean $\bm \theta$ and covariance matrix $\bm{\Sigma}_{^{\widehat{\bm \theta}}} = {\cal J}({\bm \theta})^{-1}$, that is, as $n \to \infty$, we have that
\begin{equation*}%\label{eq:fisherapp}
\sqrt{n}(\widehat{{\bm\theta}}-{\bm\theta}) \; \stackrel{\textrm{\scriptsize d}}{\to}\; \textrm{N}_{p+1}\left(\bm{0}_{p+1}, {\cal J}({\bm \theta})^{-1}\right),
\end{equation*}
where  $\stackrel{\textrm{\scriptsize d}}\to$ means ``convergence in distribution to", ${\cal J}({\bm \theta}) = \lim \limits_{n\to\infty}(1/n){\cal I}({\bm \theta})$, with ${\cal I}({\bm\theta})$ being the expected Fisher information matrix. Notice that $\widehat{{\cal I}}({\bm \theta})^{-1}$ is a consistent estimator of the asymptotic variance-covariance matrix of $\widehat{\bm \theta}$. However, in practice, we may approximate the expected Fisher information matrix by its observed version \citep{eh:78}, which can be obtained from the Hessian matrix. Furthermore, the corresponding standard errors (SEs) 
may be approximated by using the diagonal elements of its inverse. The corresponding Hessian matrix is given by 
\begin{equation*}%\label{eq:hessian}
 \ddot{\bm \ell} = \left( \begin{array}{cccc}
\textrm{tr}(\bm{G}) & \bm{k}_{(1)}^{\top}\bm{x}_{(1)} & \bm{k}_{(2)}^{\top}\bm{x}_{(2)}  \\
\bm{x}_{(1)}^{\top}\bm{k}{(1)} & \bm{x}_{(1)}^{\top}\bm{V}_{{(1)}}\bm{x}_{(1)} & \bm{x}_{(1)}^{\top}\bm{D}\bm{x}_{(2)}  \\
\bm{x}_{(2)}^{\top}\bm{k}{(2)} & \bm{x}_{(2)}^{\top}\bm{D}\bm{x}_{(1)}  & \bm{x}_{(2)}^{\top}\bm{V}_{(2)}\bm{x}_{(2)} \end{array} \right),
 \end{equation*}
where 
\begin{eqnarray*}
\bm{V}_{(1)}&=& \textrm{diag} \{v_{\textrm{\tiny{(1)1}}}({\bm \theta}),v_{\textrm{\tiny{(1)2}}}({\bm \theta}),v_{\textrm{\tiny{(1)3}}}({\bm \theta}),\ldots,v_{\textrm{\tiny{(1)n}}}({\bm \theta})\},\\
\bm{V}_{(2)}&=& \textrm{diag} \{v_{\textrm{\tiny{(2)1}}}({\bm \theta}), v_{\textrm{\tiny{(2)2}}}({\bm \theta}), v_{\textrm{\tiny{(2)3}}}({\bm \theta}),\ldots,v_{\textrm{\tiny{(2)n}}} ({\bm \theta})\},\\
\bm{k}{(1)}&=&(k_{\textrm{\tiny{(1)1}}}({\bm \theta}),k_{\textrm{\tiny{(1)2}}}({\bm \theta}), k_{\textrm{\tiny{(1)3}}}({\bm \theta}),\ldots,k_{\textrm{\tiny{(1)n}}}({\bm \theta}))^{\top},\\
\bm{k}{(2)}&=&(k_{\textrm{\tiny{(2)1}}}({\bm \theta}),k_{\textrm{\tiny{(2)2}}}({\bm \theta}), k_{\textrm{\tiny{(2)3}}}({\bm \theta}),\ldots,k_{\textrm{\tiny{(2)n}}}({\bm \theta}))^{\top},\\
\bm{D}&=& \textrm{diag} \{d_{1}({\bm \theta}),d_{2}({\bm \theta}), d_{3}({\bm \theta}),\ldots,d_{n}({\bm \theta})\},\\\bm{G}& =&\textrm{diag}\{g_{1}({\bm \theta}), g_{2}({\bm \theta}), g_{3}({\bm \theta}),\ldots,g_{n}({\bm \theta})\},\end{eqnarray*}
with
\begin{align*}
\begin{split}
g_{i}({\bm\theta}) &=
\begin{cases}
\frac{1}{\alpha^2}\left(\frac{\exp\left(\bm x^\top_{(2)_i}\bm\beta_{(2)}\right)\phi(\zeta_{2_i}^{\textrm{\tiny{c}}})\zeta_{1_i}^{\textrm{\tiny{c}}}}{{\left(1+\exp\left(\bm x^\top_{(2)_i}\bm\beta_{(2)}\right)\Phi(\zeta_{2_i}^{\textrm{\tiny{c}}})\right)}} + \frac{\zeta_{2_i}^{2\textrm{\tiny{c}}}\phi(\zeta_{2_i}^{\textrm{\tiny{c}}})\exp\left(\bm x^\top_{(2)_i}\bm\beta_{(2)}\right)}{\left(1+\exp\left(\bm x^\top_{(2)_i}\bm\beta_{(2)}\right)\Phi(\zeta_{2_i}^{\textrm{\tiny{c}}})\right)}\right)  \\ \nonumber -\frac{1}{\alpha^2}\left(\frac{\phi^{2}(\zeta_{2_i}^{\textrm{\tiny{c}}})\zeta_{2_i}^{2\textrm{\tiny{c}}}\exp\left(2\bm x^\top_{(2)_i}\bm\beta_{(2)}\right)}{\left(1+\exp\left(\bm x^\top_{(2)_i}\bm\beta_{(2)}\right)\Phi(\zeta_{2_i}^{\textrm{\tiny{c}}})\right)}\right), & \, i=1,\ldots,m;\\
-\frac{1}{\alpha^2}(3\zeta_{2_i}^{2}-1), & \, i=m+1,\ldots,n;\\
\end{cases}
\end{split}
\end{align*}
\begin{align*}
\begin{split}
k_{(1)_i}({\bm\theta}) &=
\begin{cases}
\frac{\exp\left(\bm x^\top_{(2)_i}\bm\beta_{(2)}\right)}{2\alpha}
\left(\frac{\zeta_{1_i}^{\textrm{\tiny{c}}}\phi\left(\zeta_{2_i}^{\textrm{\tiny{c}}}\right)+\zeta_{1_i}^{\textrm{\tiny{c}}}\phi\left(\zeta_{2_i}^{\textrm{\tiny{c}}}\right)\zeta_{2_i}^{2\textrm{\tiny{c}}}}{\left(1+\exp\left(\bm x^\top_{(2)_i}\bm\beta_{(2)}\right)\left(\Phi\left(\zeta_{2_i}^{\textrm{\tiny{c}}}\right)-1\right)\right)} + \frac{\exp\left(\bm x^\top_{(2)_i}\bm\beta_{(2)}\right)\phi^{2}\left(\zeta_{2_i}^{\textrm{\tiny{c}}}\right)\zeta_{1_i}^{\textrm{\tiny{c}}}\zeta_{2_i}^{\textrm{\tiny{c}}}}{\left(1+\exp\left(\bm x^\top_{(2)_i}\bm\beta_{(2)}\right)\Phi(\zeta_{2_i}^{\textrm{\tiny{c}}})\right)^{2}}\right),
& \, i=1,\ldots,m;\\
\frac{1}{\alpha}(\zeta_{1_i}\zeta_{2_i}), & \, i=m+1,\ldots,n;\\
\end{cases}
\end{split}\\
\begin{split}
k_{(2)_i}({\bm\theta}) &=
\begin{cases}
-\frac{\exp\left(\bm x^\top_{(2)_i}\bm\beta_{(2)}\right)\phi(\zeta_{2_i}^{\textrm{\tiny{c}}})\zeta_{2_i}^{\textrm{\tiny{c}}}}{\alpha}\left(\frac{{\left(1+\exp\left(\bm x^\top_{(2)_i}\bm\beta_{(2)}\right)\Phi(\zeta_{2_i}^{\textrm{\tiny{c}}})\right)} - 1}{{\left(1+\exp\left(\bm x^\top_{(2)_i}\bm\beta_{(2)}\right)\Phi(\zeta_{2_i}^{\textrm{\tiny{c}}})\right)}^{2}}\right), & \, i=1,\ldots,m;\\
0, & \, i=m+1,\ldots,n;
\end{cases}
\end{split} 
\end{align*}
\begin{align*}
\begin{split}
d_{i}({\bm\theta}) &=
\begin{cases}
-\frac{1}{2}\left(\frac{\exp\left(\bm x^\top_{(2)_i}\bm\beta_{(2)}\right)\phi(\zeta_{2_i}^{\textrm{\tiny{c}}})\zeta_{1_i}^{\textrm{\tiny{c}}}}{{\left(1+\exp\left(\bm x^\top_{(2)_i}\bm\beta_{(2)}\right)\Phi(\zeta_{2_i}^{\textrm{\tiny{c}}})\right)}} - \frac{\exp\left(2 \bm x^\top_{(2)_i}\bm\beta_{(2)}\right)\phi(\zeta_{2_i}^{\textrm{\tiny{c}}})\zeta_{1_i}^{\textrm{\tiny{c}}}\Phi(\zeta_{2_i}^{\textrm{\tiny{c}}})}{{\left(1+\exp\left(\bm x^\top_{(2)_i}\bm\beta_{(2)}\right)\Phi(\zeta_{2_i}^{\textrm{\tiny{c}}})\right)}^{2}}\right), & \, i=1,\ldots,m;\\
0, & \, i=m+1,\ldots,n;
\end{cases}
\end{split}
\end{align*}
\begin{align*}
\begin{split}
v_{(1)_i}({\bm\theta}) &=
\begin{cases}
\frac{\exp\left(\bm x^\top_{(2)_i}\bm\beta_{(2)}\right)}{4}\left(\frac{-\phi(\zeta_{2_i}^{\textrm{\tiny{c}}})\zeta_{2_i}^{\textrm{\tiny{c}}}+\zeta_{1_i}^{2\textrm{\tiny{c}}}\zeta_{2_i}^{\textrm{\tiny{c}}}\phi(\zeta_{2_i}^{\textrm{\tiny{c}}})}{{\left(1+\exp\left(\bm x^\top_{(2)_i}\bm\beta_{(2)}\right)\Phi(\zeta_{2_i}^{\textrm{\tiny{c}}})\right)}} + \frac{\phi^{2}(\zeta_{2_i}^{\textrm{\tiny{c}}})\zeta_{1_i}^{2\textrm{\tiny{c}}}\exp\left(\bm x^\top_{(2)_i}\bm\beta_{(2)}\right)}{\left(1+\exp\left(\bm x^\top_{(2)_i}\bm\beta_{(2)}\right)\Phi(\zeta_{2_i}^{\textrm{\tiny{c}}})\right)^{2}}\right),  & \, i=1,\ldots,m;\\
\frac{1}{4}(1-(\zeta_{2_i}^{2}/\zeta_{1_i}^{2}) - \zeta_{1_i}^{2}-\zeta_{2_i}^{2}), & \, i=m+1,\ldots,n,
\end{cases}
\end{split}\\
\begin{split}
v_{(2)_i}({\bm\theta}) &=
\begin{cases}
\frac{\exp\left({\bm x}_{(2)_i}^{\top}{\bm\beta}_{(2)}\right)\left(\Phi\left(\zeta_{2_i}^{\textrm{\tiny{c}}}\right)-1\right)}{\left(1+ \exp({\bm x}_{(2)_i}^{\top}{\bm\beta}_{(2)})\left(\Phi\left(\zeta_{2_i}^{\textrm{\tiny{c}}}\right)-1\right)\right)}- \frac{\left({\exp\left({\bm x}_{(2)_i}^{\top}{\bm\beta}_{(2)}\right)\left(\Phi\left(\zeta_{2_i}^{\textrm{\tiny{c}}}\right)-1\right)}\right)^{2}}{{\left(1 + \exp\left({\bm x}_{(2)_i}^{\top}{\bm\beta}_{(2)}\right)\left(\Phi\left(\zeta_{2_i}^{\textrm{\tiny{c}}}\right)-1\right)\right)}^{2}}- \tau_{i} + \tau_{i}^{2},  & \, i=1,\ldots,m;\\
\tau_{i} - \tau_{i}^{2}, & \, i=m+1,\ldots,n.
\end{cases}
\end{split}
\end{align*}

\subsection{Residual analysis}\label{sec:3.4}

We consider the GCS residual to assess goodness of fit and departures from the assumptions of the model. This residual is  often used in generalized linear models and survival analysis.  The GCS residual is given by 
\begin{equation*}
%\label{resCS}
r^{ \textrm{\tiny GCS}}_i=-\log\big( \widehat{S}(y_i;\widehat{\bm\theta}) \big),  \quad i=1,\ldots,n,
\end{equation*}
where $S$ is the corresponding survival function fitted to the data. If the model is correctly specified, then the GCS residual has a unit  exponential distribution, EXP(1) in short.  

\section{Numerical studies} \label{sec:4}

In this section, we provide the numerical results of our study. First, we evaluate the performance of the new mixture model through MC simulations. Then, an illustration of this model is presented with the biometrical data related to the case study of Section \ref{sec:motiexa}. 

\subsection{Simulation study} %\label{sec:4}

We present an MC simulation study with 5000 replications that intends to reveal the performance of the ML estimators for 
the parameters of the Bernoulli/BS model. The sample sizes considered are $n=100, 300, 500$, with 
parameters $\alpha = 0.1, 0.5, 1, 2, 4, \bm \beta_{(1)}=(0.2,0.5)^{\top}$ and $ \bm \beta_{(2)}=(1,2)^{\top}$. 
We consider one covariate $\bm X$, where $X \sim \textrm{Uniform}(0,1)$. The generated values for the response variable were obtained as follows
\begin{eqnarray*}
T_{i}=
\begin{cases}
1, & \textrm{with probability} \;\; 1-\frac{\exp\left(\beta_{\textrm{\tiny{(2)0}}} + \beta_{\textrm{\tiny{(2)1}}}x_{i}\right)}{1+\exp\left(\beta_{\textrm{\tiny{(2)0}}} + \beta_{\textrm{\tiny{(2)1}}}x_{i}\right)}, \\
\exp\left(\beta_{\textrm{\tiny{(1)0}}}+\beta_{\textrm{\tiny{(1)1}}}x_{i}\right)\delta_{i}, & \textrm{with probability} \;\;  \frac{\exp\left(\beta_{\textrm{\tiny{(2)0}}} + \beta_{\textrm{\tiny{(2)1}}}x_{i}\right)}{1+\exp\left(\beta_{\textrm{\tiny{(2)0}}} + \beta_{\textrm{\tiny{(2)1}}}x_{i}\right)},
\end{cases}
\end{eqnarray*}
where $\delta_i \sim \textrm{BS}(\alpha,1)$. In order to obtain $Y_{i}$ we take the natural logarithm of $T_{i}$.    

We compute the empirical mean, bias and mean squared error (MSE) in order to evaluate the performances of the estimators. All numerical evaluations were done in the R software; see \cite{R:15}. Table \ref{table:MC} presents the ML estimation results obtained for the mentioned values of sample sizes and parameters. This table allows to conclude that, for $\alpha = 0.1, 0.5, 1.0, 2.0$, as the sample size increases, the bias and MSE of all estimators decrease, as expected. In general, the results show the good performance of the ML estimators of the Bernoulli/BS model parameters.

\begin{table}[!ht]
\small
\centering
\renewcommand{\arraystretch}{0.8}
\renewcommand{\tabcolsep}{0.2cm}
\caption{Summary statistics from simulated Bernoulli/BS data for the indicated estimator and sample size.}
\label{table:MC}
\begin{adjustbox}{max width=\textwidth}
  \begin{tabular}{llrrrrrrrrrrrr} 
\toprule 
\multirow{2}*{$\alpha$} &\multirow{2}*{$\bm \theta$} & & \multicolumn{3}{c}{$n=100$} && \multicolumn{3}{c}{$n =300$}&& \multicolumn{3}{c}{$n=500$}  \\  \cline{4-6}  \cline{8-10} \cline{12-14} \\ \vspace{0.1cm} 
  & & & \multicolumn{1}{c}{Mean} & \multicolumn{1}{c}{Bias} & \multicolumn{1}{c}{MSE}  & & \multicolumn{1}{c}{Mean} & \multicolumn{1}{c}{Bias}  & \multicolumn{1}{c}{MSE} & &   \multicolumn{1}{c}{Mean}  & \multicolumn{1}{c}{Bias}  & \multicolumn{1}{c}{MSE}   \\ \midrule
0.1 &  $\alpha$ && 0.21103 & 0.11103 & 0.62099 && 0.14891 & 0.04891 & 0.00328 && 0.10005 & 0.00005 & 0.00039  \\
    &  $\beta_{\textrm{\tiny{(1)0}}}=0.2$ && 0.21215  & 0.01215 & 1.03058 && 0.20326 & 0.00326 & 0.08643  && 0.20309 & 0.00309 & 0.01018 \\
    &  $\beta_{\textrm{\tiny{(1)1}}}=0.5$  && 0.48652  & $-$0.01348 & 0.98195 && 0.50374 & 0.00374 & 0.09117 && 0.50057 &  0.00057 & 0.00977  \\
    &  $\beta_{\textrm{\tiny{(2)0}}}=1$  && 1.00140  & 0.00140 & 1.17457 && 1.00115 & 0.00115 & 0.08867 && 0.99797 & $-$0.00203 & 0.00997 \\\vspace{0.1cm}
    &  $\beta_{\textrm{\tiny{(2)1}}}=2$  && 1.98399  & $-$0.01600 & 1.04944 && 2.00469 & 0.00469  & 0.09159 && 2.00104 & 0.00104 &  0.01007 \\ \midrule

0.5 &  $\alpha$ && 0.50211 & 0.00211 & 0.01549 && 0.49848 & $-$0.00152 & 0.00343 && 0.50004 & 0.00004 & 0.00032 \\
    &  $\beta_{\textrm{\tiny{(1)0}}}=0.2$ && 0.22839 & 0.02839 & 0.81110 && 0.21216 & 0.01216 & 0.14482 &&  0.20293 & 0.00293 & 0.00919   \\
    &  $\beta_{\textrm{\tiny{(1)1}}}=0.5$  && 0.49694 & $-$0.00306 & 0.79049 && 0.50993 & 0.00993  & 0.15961 && 0.50054  & 0.00054  & 0.00881  \\
    &  $\beta_{\textrm{\tiny{(2)0}}}=1$  && 1.00613 & 0.00613 & 0.83217 && 1.01332 & 0.01332 & 0.17506 && 0.99807 & $-$0.00192  & 0.00899 \\\vspace{0.1cm}
    &  $\beta_{\textrm{\tiny{(2)1}}}=2$  && 1.98798 & $-$0.01201 & 0.81395 && 2.01255 & 0.01255 & 0.16855 && 2.00098 & 0.00098 & 0.00908 \\ \midrule
		
1 &  $\alpha$ && 0.99855 & $-$0.00145 & 0.06463 && 0.99941 & $-$0.00059 & 0.00279 && 1.00004 & 0.00005 & 0.00039  \\
    &  $\beta_{\textrm{\tiny{(1)0}}}=0.2$ && 0.30713 & 0.10713 & 0.46320 && 0.20529 & 0.00529 & 0.03719 && 0.20278 & 0.00278 & 0.00825 \\
    &  $\beta_{\textrm{\tiny{(1)1}}}=0.5$  && 0.54346 & 0.04346 & 0.46502 && 0.50423 & 0.00423 & 0.03999 && 0.50052 & 0.00052 & 0.00791  \\
    &  $\beta_{\textrm{\tiny{(2)0}}}=1$  && 1.09830 & 0.09830 & 0.64609 && 1.00368 & 0.00368 & 0.04167 && 0.99817 & $-$0.00183 & 0.00807 \\\vspace{0.1cm}
    &  $\beta_{\textrm{\tiny{(2)1}}}=2$  && 2.04050 & 0.04050 & 0.54142 && 2.00462 & 0.00462 & 0.04132 && 2.00093 & 0.00093 & 0.00815 \\ \midrule
		
2 &  $\alpha$ && 2.01445 & 0.01445 & 0.06145 && 1.99856 & $-$0.00144 & 0.00159 && 2.00002 & 0.00002 & 0.00010  \\ 
    &  $\beta_{\textrm{\tiny{(1)0}}}=0.2$ && 0.35064 & 0.15064 & 0.27488 && 0.20109 & 0.00109 & 0.00959 && 0.20030 & 0.00030 & 0.00010  \\
    &  $\beta_{\textrm{\tiny{(1)1}}}=0.5$  && 0.56994 & 0.06994 & 0.24613 && 0.50125 & 0.00125 & 0.01012 && 0.50057 & 0.00006 & 0.00010  \\
    &  $\beta_{\textrm{\tiny{(2)0}}}=1$  && 1.20834 & 0.20834 & 0.72936 && 1.00040 & 0.00040 & 0.00985 && 0.99979 & $-$0.00021 & 0.00010 \\ \vspace{0.1cm}
    &  $\beta_{\textrm{\tiny{(2)1}}}=2$  && 2.10056 & 0.10056 & 0.38308 && 2.00157 & 0.00157 & 0.01018 && 2.00010 & 0.00010 & 0.00010  \\  \midrule
		
4 &  $\alpha$ &&  5.74480 & 1.74480 & 31.44706 && 5.63512 & 1.63512 & 18.12119 && 4.21585 & 0.21585 & 4.01609 \\
    &  $\beta_{\textrm{\tiny{(1)0}}}=0.2$ && 0.84559 & 0.64559 & 3.61015 && 0.90587 & 0.70587 & 3.46159 && 0.26577 & 0.06577 & 0.36885 \\
    &  $\beta_{\textrm{\tiny{(1)1}}}=0.5$  && 0.77905 & 0.27905 & 0.90060 && 0.91386 & 0.41386 & 1.16450 && 0.53409 & 0.03409 & 0.10023 \\
    &  $\beta_{\textrm{\tiny{(2)0}}}=1$  && 3.18593 & 2.8593 & 40.9525 && 3.29328 & 2.29328 & 34.75177 && 1.22581 & 0.22581 & 4.40974 \\
    &  $\beta_{\textrm{\tiny{(2)1}}}=2$  && 3.09786 & 1.09786 & 10.54635 && 3.22245 & 1.22245 & 9.94170 && 2.10635 & 0.10635 & 0.97295 \\
 \bottomrule 
\end{tabular}
\end{adjustbox}
\end{table}

\subsection{Illustrative example}%\label{example:01}

\begin{figure}[!ht]
\centering
\psfrag{R}[c]{\scriptsize{empirical quantile}}
\psfrag{Q}[c]{\scriptsize{theoretical quantile}}
\psfrag{0}[c][c]{\scriptsize{0}}
\psfrag{1}[c][c]{\scriptsize{1}}
\psfrag{2}[c][c]{\scriptsize{2}}
\psfrag{3}[c][c]{\scriptsize{3}}
\psfrag{4}[c][c]{\scriptsize{4}}
\psfrag{5}[c][c]{\scriptsize{5}}
\psfrag{6}[c][c]{\scriptsize{6}}
\psfrag{7}[c][c]{\scriptsize{7}}
\psfrag{8}[c][c]{\scriptsize{8}}
\psfrag{9}[c][c]{\scriptsize{9}}
\psfrag{10}[c][c]{\scriptsize{10}}
\psfrag{15}[c][c]{\scriptsize{15}}
\psfrag{-0.002}[c][c]{\scriptsize{$-$.002}}
\psfrag{0.000}[c][c]{\scriptsize{.0}}
\psfrag{0.002}[c][c]{\scriptsize{.002}}
\psfrag{0.004}[c][c]{\scriptsize{.004}}
\psfrag{0.006}[c][c]{\scriptsize{.006}}
\psfrag{0.008}[c][c]{\scriptsize{.008}}
\psfrag{0.010}[c][c]{\scriptsize{.010}}
\psfrag{0.0}[c][c]{\scriptsize{0.0}}
\psfrag{0.1}[c][c]{\scriptsize{0.1}}
\psfrag{0.2}[c][c]{\scriptsize{0.2}}
\psfrag{0.3}[c][c]{\scriptsize{0.3}}
\psfrag{0.4}[c][c]{\scriptsize{0.4}}
\psfrag{0.5}[c][c]{\scriptsize{0.5}}
\psfrag{0.6}[c][c]{\scriptsize{0.6}}
\psfrag{0.7}[c][c]{\scriptsize{0.7}}
\psfrag{0.8}[c][c]{\scriptsize{0.8}}
\psfrag{1.0}[c][c]{\scriptsize{1.0}}
\psfrag{0}[c][c]{\scriptsize{0}}
\psfrag{50}[c][c]{\scriptsize{50}}
\psfrag{100}[c][c]{\scriptsize{100}}
\psfrag{150}[c][c]{\scriptsize{150}}
\psfrag{200}[c][c]{\scriptsize{200}}
\psfrag{250}[c][c]{\scriptsize{250}}
\psfrag{300}[c][c]{\scriptsize{300}}
\psfrag{328}[c]{\scriptsize{$328$}}
\psfrag{330}[l]{\scriptsize{$330$}}
\psfrag{in}[c]{\scriptsize{index}}
\psfrag{GCDn}{\scriptsize{GCD($\bm\theta$)}}
\subfigure[tobit-BS]{\includegraphics[height=7.5cm,width=7.5cm]{Envelope-BS.eps}}
\subfigure[Bernoulli/BS]{\includegraphics[height=7.5cm,width=7.5cm]{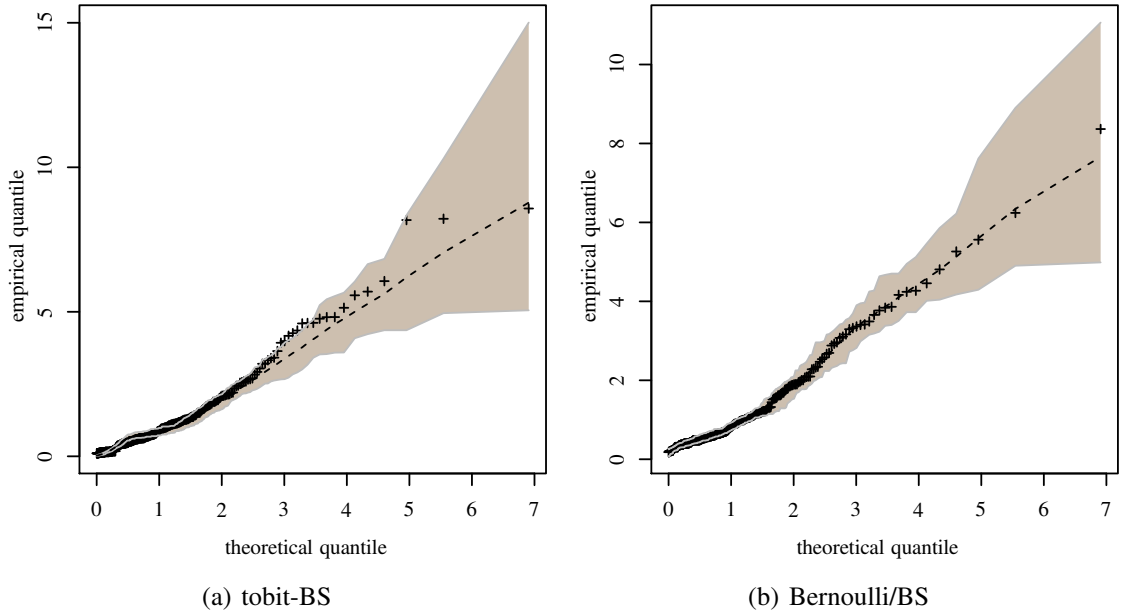}}
\caption{\small QQ plot and its envelope for the GCS residual with the indicated model using vaccine data.}%
\label{fig:qqplots}
\end{figure}

We illustrate the proposed methodology by applying it to the real-world biometry data set described in Section \ref{sec:motiexa}. Here, we present the estimation and checking results for the proposed Bernoulli/BS model with these data. For comparison, the results of the standard tobit, in addition to the tobit-BS model, are given as well. The Bernoulli/BS model has a logit link function with the same covariates used in the continuous component. Table \ref{tab:results} shows the ML estimates, computed by the BFGS method,  SEs, $p$-values of the $t$-test and the Akaike (AIC) information (BIC) criterion. From this table, note that the Bernoulli/BS model provides better adjustment compared to the other models based on the value of AIC. Figure~\ref{fig:qqplots} displays the QQ plots with simulated envelope of the GCS residual. This figure shows that the GCS residuals provide an excellent agreement with the EXP(1) distribution for the Bernoulli/BS models. 
\begin{table}[!ht]
\small
\centering
\renewcommand{\arraystretch}{0.8}
\renewcommand{\tabcolsep}{0.2cm}
\caption{ML estimates (with SE in parentheses) and AIC values for the indicated models with vaccine data.}
\label{tab:results}
\begin{adjustbox}{max width=\textwidth}
 \begin{tabular}{lccccccccccccc} \toprule
\multirow{2}*{ Model } &\multirow{2}*{AIC}  & & \multicolumn{4}{c}{Logit component} & & \multicolumn{5}{c}{Continuous component} \\ 
\cline{4-7}  \cline{9-13} \\ 
             &        & & Constant        & EZ          & HI      & FEM     &  & $\alpha$   & Constant            & EZ        & HI         & FEM \\ \midrule
Tobit         & 1299.27& &            &             &         &         &  &  2.573      & 0.597{**}      & 0.225     & $-$0.228   & 0.271 \\ \vspace{0.2cm}
             &        & &            &             &         &         &  &  (0.047)   & (0.288)        & (0.297)   & (0.295)    & (0.296) \\
Tobit-BS      & 1168.60& &            &             &         &         &  &  1.545     &  $-$0.910{***} & 0.188{*}  & 0.074      & 0.121 \\ \vspace{0.2cm}
             &        & &            &             &         &         &  &  (0.048)   & (0.105)        & (0.111)   & (0.109)    & (0.110) \\   			
Bernoulli/BS  & 1085.32& &0.762{***}  &0.739{***}   & 0.347   &$-$0.269 &  &  1.208     & $-$0.061       &$-$0.159   & $-$0.180   &  0.284{**} \\ 
             &        & & (0.245)    & (0.282)     & (0.270) & (0.271) &  &  (0.064)   & (0.136)        & (0.143)   & (0.143)    & (0.144)\\ 
\bottomrule \vspace{-0.3cm} \\ 
\multicolumn{12}{l}{\small{*10\% of significance,** 5\% of significance and ***1\% of significance.}} \\
\end{tabular}
\end{adjustbox}
\end{table}

We note that, in the fitted Bernoulli/BS model presented in Table \ref{tab:results}, only the variable FEM is significant for the continuous component, whereas for the logit component only the variable EZ was significant. Therefore, for the Bernoulli/BS model, the fitted final model is given by 
$$
\widehat{\pi}_i = \frac{1}{1+\exp(\underset{(\mathbf{0.162})}{0.762} + \underset{(\mathbf{0.260})}{0.657} \times \textrm{EZ})} \quad \textrm{and} \quad \widehat{\mu}_{i} = -\underset{(\mathbf{0.089})}{0.146} + \underset{(\mathbf{0.129})}{0.233} \times \textrm{FEM},
$$
with $\widehat{\alpha} =  1.166\,({0.053})$. A glance at these results indicate that, receipt of Edmonston-Zagreb strain vaccine is related with an increase 
$\exp(0.657)= 1.193$ in the odds ratio of being above the detection limit. Moreover, the Bernoulli/BS model suggests that girls have $\exp(0.233)= 1.263$ greater concentration of measles antibody 
than boys. 

\section{Concluding remarks and future research}\label{sec:6}

We have introduced a new continuous-discrete mixture fixed-effect model whose continuous part follows a Birnbaum-Saunders distribution and its discrete-part a Bernoulli distribution. This model is very flexible and useful for highly censored data. Our investigation was based on a biometrical case-study related to measles vaccines in Haiti. We have performed estimation and inference based on the maximum likelihood method. A Monte Carlo simulation study has shown the good performance of the maximum likelihood estimators. The numerical results of the case-study have proved the excellent agreement between the Bernoulli/Birnbaum-Saunders model and the data, improving the fitting in relation to other competitors as the standard tobit and tobit-BS models. 

As part of further research, it is of interest to discuss influence diagnostic tools with more detail and depth to detect globally and locally influential cases. This will allow us to evaluate changes in the model's significance and consequently in the medical decisions. In addition, multivariate models can also be explored. Work on some of these issues is currently in progress and we hope to report some findings in future papers.

%\paragraph{Acknowledgments}
%This study was partially supported by CAPES from the Brazilian government and FONDECYT 1160868 project grants from the Chilean government.

\small

%\bibliographystyle{apalike}
%\bibliography{bibliography}

\end{document}